\documentclass[twocolumn,twocolappendix]{aastex63}

\usepackage{comment}
\usepackage{dcolumn}
\usepackage[version=3]{mhchem}
\usepackage{amssymb}
\let\tablenum\relax
\usepackage{siunitx}
\usepackage{threeparttable}
\usepackage{color}

\newcommand{\rev}[1]{{{#1}}}
\newcommand{\revrev}[1]{{{#1}}}

\definecolor{red}{rgb}{0.8,0.0,0.0}
\definecolor{blue}{rgb}{0.0,0.0,0.8}
\definecolor{green}{rgb}{0.0,0.5,0.0}

\received{September 29, 2021}
\accepted{January 23, 2022}

\submitjournal{ApJS}

\shortauthors{Y. Kasagi et al.}
\graphicspath{{./}{figures}}

\begin{document}

\title{Dippers from TESS Full-frame Images. II. Spectroscopic Characterization of Four Young Dippers}

\correspondingauthor{Yui Kasagi}
\email{yui.kasagi@grad.nao.ac.jp}

\author[0000-0002-8607-358X]{Yui Kasagi}
\affiliation{Department of Astronomy, School of Science, The Graduate University for Advanced Studies, SOKENDAI, \\ 2-21-1 Osawa, Mitaka, Tokyo, Japan}

\author[0000-0001-6181-3142]{Takayuki Kotani}
\affiliation{Department of Astronomy, School of Science, The Graduate University for Advanced Studies, SOKENDAI, \\ 2-21-1 Osawa, Mitaka, Tokyo, Japan}
\affiliation{Astrobiology Center, 2-21-1 Osawa, Mitaka, Tokyo 181-8588, Japan}
\affiliation{National Astronomical Observatory of Japan, 2-21-1 Osawa, Mitaka, Tokyo 181-8588, Japan}

\author[0000-0003-3309-9134]{Hajime Kawahara}
\affiliation{Department of Earth and Planetary Science, The University of Tokyo, 7-3-1, Hongo, Tokyo, Japan}
\affiliation{Research Center for the Early Universe, School of Science, The University of Tokyo, Tokyo 113-0033, Japan}

\author{Tomoyuki Tajiri}
\affiliation{Department of Earth and Planetary Science, The University of Tokyo, 7-3-1, Hongo, Tokyo, Japan}

\author{Takayuki Muto}
\affiliation{Division of Liberal Arts, Kogakuin University, 1-24-2 Nishi-Shinjyuku, Shinjyuku-ku, Tokyo 163-8677, Japan}

\author[0000-0001-8877-4497]{Masataka Aizawa}
\affiliation{Tsung-Dao Lee Institute, Shanghai Jiao Tong University, 800 Dongchuan Road, Shanghai 200240, China}

\author{Michiko S. Fujii}
\affiliation{Department of Astronomy, The University of Tokyo, 7-3-1, Hongo, Tokyo, Japan}

\author[0000-0001-6924-8862]{Kohei Hattori}
\affiliation{National Astronomical Observatory of Japan, 2-21-1 Osawa, Mitaka, Tokyo 181-8588, Japan}
\affiliation{Institute of Statistical Mathematics, 10-3 Midoricho, Tachikawa, Tokyo 190-0014, Japan}

\author[0000-0003-1298-9699]{Kento Masuda}
\affiliation{Department of Earth and Space Science, Osaka University, Osaka 560-0043, Japan}

\author[0000-0002-3001-0897]{Munetake Momose}
\affiliation{College of Science, Ibaraki University, 2-1-1 Bunkyo, Mito, Ibaraki 310-8512, Japan}

\author[0000-0001-5797-6010]{Ryou Ohsawa}
\affiliation{Institute of Astronomy, The University of Tokyo, 2-21-1 Osawa, Mitaka, Tokyo 181-0015, Japan}

\author{Satoshi Takita}
\affiliation{Institute of Astronomy, The University of Tokyo, 2-21-1 Osawa, Mitaka, Tokyo 181-0015, Japan}

\begin{abstract}

Photometric monitoring by the Transiting Exoplanet Survey Satellite (TESS) has discovered not only periodic signals by transiting exoplanets but also episodic or quasi-periodic dimming around young stellar objects.
The mechanisms of the dimming of these objects, so-called ``dippers'',  are thought to be related to the property of the accretion or the structure of protoplanetary disks especially in regions close to the host star.
Recently, we have created the catalog of dippers from the one year of TESS Full Frame Image (FFI) data.
In this paper, we report spectral features of four newly found dippers in that catalog and show that they potentially shed light on the dimming mechanisms.
We found that all of the targets exhibit the \ce{H\alpha} emission line, which is an indicator of an ongoing accretion.
Based on its line profiles and/or their variability, we characterized the properties of the disks of each source, which can support the dimming mechanisms due to a dusty disk wind or an accretion warp.
Also, we found an interesting dipper (TIC 317873721), ``close-in binary dipper,'' showing the complex variability of the line profile and the large radial velocity variation.
Since the dimming intervals are similar to the orbital period of the binary, we suggest that the dips are caused by dust in the warp of accretion from a circumbinary disk onto stars.
Such a close-in ($<$ \SI{0.1}{au}) binary dipper is rarely reported so far, further investigation will reveal the new aspect of the disk evolution and planetary formation.

\end{abstract}

\keywords{Variable stars (1761), Protoplanetary disks (1300)}

\section{Introduction} 
\label{sec:intro}

Observations of Young Stellar Objects (YSOs) provide an important clue to study planet formation processes \citep[c.f. ][]{Hillenbrand2008}.
Some YSOs show episodic/quasi-periodic dimmings in their light curves with a duration of $\sim$\SI{1}{day}. 
These irregularly dimming stars are called ``dippers,'' which are distinct from any stars showing regular variabilities. 
So far, the detailed mechanisms for such variability are not well understood. 
At first, it was thought that dippers were caused by their circumstellar environment, such as accretion from a warped inner disk \citep{Bouvier1999}. 
Also, various mechanisms have been considered thus far including the transiting circumstellar clumps \citep{Ansdell2016}. 
In this scenario, the dippers are intriguing objects to understand the dynamic process of disk evolution and planet formation. 
However, we need to reconsider such a classical view on the dippers owing to the recent long-term, high-cadence monitoring campaigns of young stars (e.g., Kepler, TESS).

The survey of the dippers has so far been limited mostly in star-forming regions due to small field-of-views (FOVs) of K2 and CoRoT, and also due to constraints on the pointing for K2.  
Transiting Exoplanet Survey Satellite (TESS), launched in 2018, has drastically changed the situation.  
TESS is an all-sky monitoring satellite with extremely precise photometry.
It provides two types of data: short cadence (\SI{2}{min}) time-series of preselected targets and the full-frame images (FFI) of \SI{30}{min} cadence.
FFI is close to raw data, providing us with a large discovery space for new types of stellar variability.

Recently, we have developed a pipeline to extract clean light curves from FFIs using recent machine learning technologies including the convolutional neural network \citep[hereafter ``Paper I'']{Tajiri2020}.  
We created $4\times10^{6}$ light curves from TESS FFIs of Sectors 1-13, which corresponds to the first year of the TESS mission that covers almost half of the hemisphere.   
Extensive investigation of TESS images leads to the discovery of dipper-like phenomena for 35 stars across the sky, which is currently the largest, homogeneous catalog of dippers with episodic/quasi-periodic dimming with infrared excess. 
This catalog includes TIC 284730577, which is identified as a star showing a dipper-like light curve by \citet{Gaidos2019} as well.
This confirms that the newly developed pipeline indeed can identify dipper-like photometric variation.

Our catalog includes several interesting examples of dippers that do not fit the classical view of the dippers. 
For example, in Paper I, we discussed the detailed property of TIC 43488669, one of the `runaway' dippers which exhibited large peculiar velocity ($>$\SI{30}{km.s^{-1}}) and was not associated with any circumstellar materials. 
Another example is an isolated YSO, which is isolated from any molecular clouds and clusters but does not exhibit large peculiar motion. 
Because these dippers are considered to have different origins or evolution processes from other dippers, they will significantly expand the picture of the dipper.
In this paper, we show the results of spectroscopic observations for some unexpected classes of dippers and some classical-type dippers in our catalog to characterize their disk properties and evolutional phase.

In Section \ref{sec:observations}, we describe the spectroscopic observations for dippers and sub-millimeter observations for two dippers.
In Section \ref{sec:results}, we show the results of spectral analysis.
In Section \ref{sec:scenarios}, we explain the estimated cause of dimming for these dippers.
The discussion and summary are in Section \ref{sec:summary}.

\section{Observations and data reduction}
\label{sec:observations}
\subsection{Target}
\begin{figure}
  \centering
  \includegraphics[width=7cm]{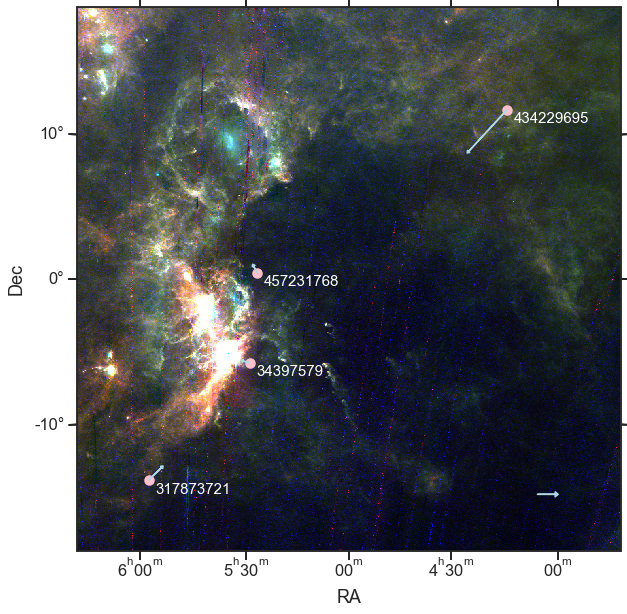}
  \caption{The distribution of the observed dippers. The length of the arrows is propotional to the tangential velocities (\si{km.s^{-1}}) calculated from Gaia DR2 data. The arrow corresponding to \SI{10}{km.s^{-1}} velocity is shown at the bottom. The background picture is synthesized from images from AKARI \citep{Takita2015}.}
  \label{fig:dippermap}
\end{figure}

We observed four of the dippers in our catalog (Table 1 of Paper I);
their TIC IDs are 457231768, 34397579, 434229695 and 317873721.
We select them because they are bright enough to be observed with high-dispersion spectrographs on telescopes in the northern hemisphere.
These dippers are located around \rev{Orion} as shown in Figure \ref{fig:dippermap}.
Their properties are listed in Table \ref{tab:targetlist}.

\begin{deluxetable*}{ccccccccccc}
\tabletypesize{\footnotesize}
\tablecaption{Properties of Targets\label{tab:targetlist}}
\tablehead{
\colhead{TIC ID$^{a}$} & \colhead{Gaia ID$^{b}$}  & \colhead{RA$^{c}$} & \colhead{Dec$^{c}$} & \colhead{T$_{\mathrm{eff}}^{c}$ (K)} & \colhead{Radius$^{c}$ (R$_{\odot}$)} 
}
\startdata
    457231768 & 3221803265361341056 & 05:27:05.47 & 00:25:07.64 &  8785 $\pm$ 158 & 1.96 $\pm$ 0.09 \\
    34397579 & 3208985468042433792 & 05:29:11.44 & -06:08:05.40 & 6330 $\pm$ 140 & 2.30 $\pm$ 0.12\\
    434229695 & 3304172564776041728 & 04:14:18.74 & 11:58:12.66 & 5312 $\pm$ 147 & 1.92 $\pm$ 0.11 \\
    317873721 & 2995720039486652672 & 05:57:53.77 & -14:05:25.29 & 8211 $\pm$ 148 & 1.67 $\pm$ 0.06\\
\enddata
\tablecomments{$^{a}$ TESS Input Catalog ID, $^{b}$ Gaia Source ID, $^{c}$ RA, Dec (J2015.5), effective temperature, and stellar radius listed in TICv8\citep{ExoFOP}}
\end{deluxetable*}  

The observations were performed several times from 2019 to 2021 with spectrometers on the Subaru telescope, the Calar Alto Observatory, the Okayama 188 cm reflector, and sub-millimeter bolometer camera on the JCMT for each target.
See Table \ref{tab:result} for the instruments used for observing each target.

\subsection{Subaru/HDS}
The observation by High-Dispersion Spectrograph  \citep[HDS:][]{Noguchi2002} installed on the Subaru telescope was performed on 16 September 2019 for all targets and 6 September 2020 for TICs 434229695 and 317873721 with the non-standard setup and a $2 \times 1$ binning without the image rotator. We use the image slicer (IS) \#$2$ with a spectral resolution of about $80,000$. The non-standard setup of the wavelength range includes H$\alpha$ and lithium \SI{6707}{\angstrom} line as an indicator of YSOs and also Mg triplet lines ($5165$--$5185$\si{\angstrom}) for estimates of stellar temperature, and radial velocity. 

The data were reduced by the standard procedure (the bias subtraction, flat fielding, order tracing/extraction, and wavelength calibration) by using IRAF, which yields one-dimensional spectra with a signal-to-noise ratio of $80$ -- $100$ per pixel at \SI{6563}{\angstrom}.

\subsection{CAHA/CAFÉ}
We performed additional observations for TICs 434229695 and 317873721 by Calar Alto Fibre-fed Échelle spectrograph  \citep[CAFÉ:][]{Aceituno2013} attached on the \SI{2.2}{m} telescope of the Calar Alto Observatory (CAHA).
We observed them on 2 and 16 October, 13 November and 18 December 2019 and on 8 November 2019 only for TIC 434229695. 
The spectral range is $407$ -- $925$ \si{nm} with spectral resolution of about $62,000$. 

The data were reduced using the CAFExtractor pipeline  \citep{Lillo-Box2020}, partly based on the CERES algorithms  \citep{Brahm2017}.
 
 \subsection{Okayama/HIDES}
 We also performed additional observation for TIC 317873721 on 20 and 25 January and 2, 4, 19, 22 and 28 February 2021 by using the High Dispersion Echelle Spectrograph  \citep[HIDES:][]{Izumiura1999} at Okayama Astrophysical Observatory (OAO).
 The observing wavelength of HIDES is $360$ -- \SI{1000}{nm} and the resolution is $52,000$ by using HE-mode.
 
The data were reduced by the standard procedure (the bias subtraction, flat fielding, order tracing/extraction, and wavelength calibration) by using IRAF.
 
 \subsection{JCMT/SCUBA-2}

We performed sub-mm photometric observations for TIC~434229695 and TIC~317873721 with Submillimeter Common-User Bolometer Array 2  \citep[SUCUBA-2:][]{Dempsey2013} on James Clerk Maxwell Telescope (JCMT) at Maunakea, Hawaii.  The observations of TIC~434229695 was conducted on 8 September 2020 UT using the Pointsource Daisy Map Mode with the observing time of 26~min.  The observations of TIC~317873721 was conducted on 3 August, 8 September, and 21 September 2020 using the Pointsource Daisy Map Mode with the observing time of 26~min.

The data were reduced using the pipeline in the Starlink software \citep{Currie2014}.  The recipe 
\verb|REDUCE_SCAN_FAINT_POINT_SOURCES|
is used for point source detection.

\section{Results}
\label{sec:results}
\subsection{Classification of TESS Light Curves}
\label{sec:LCs}
\begin{figure*}
  \centering
  \includegraphics[width=17cm]{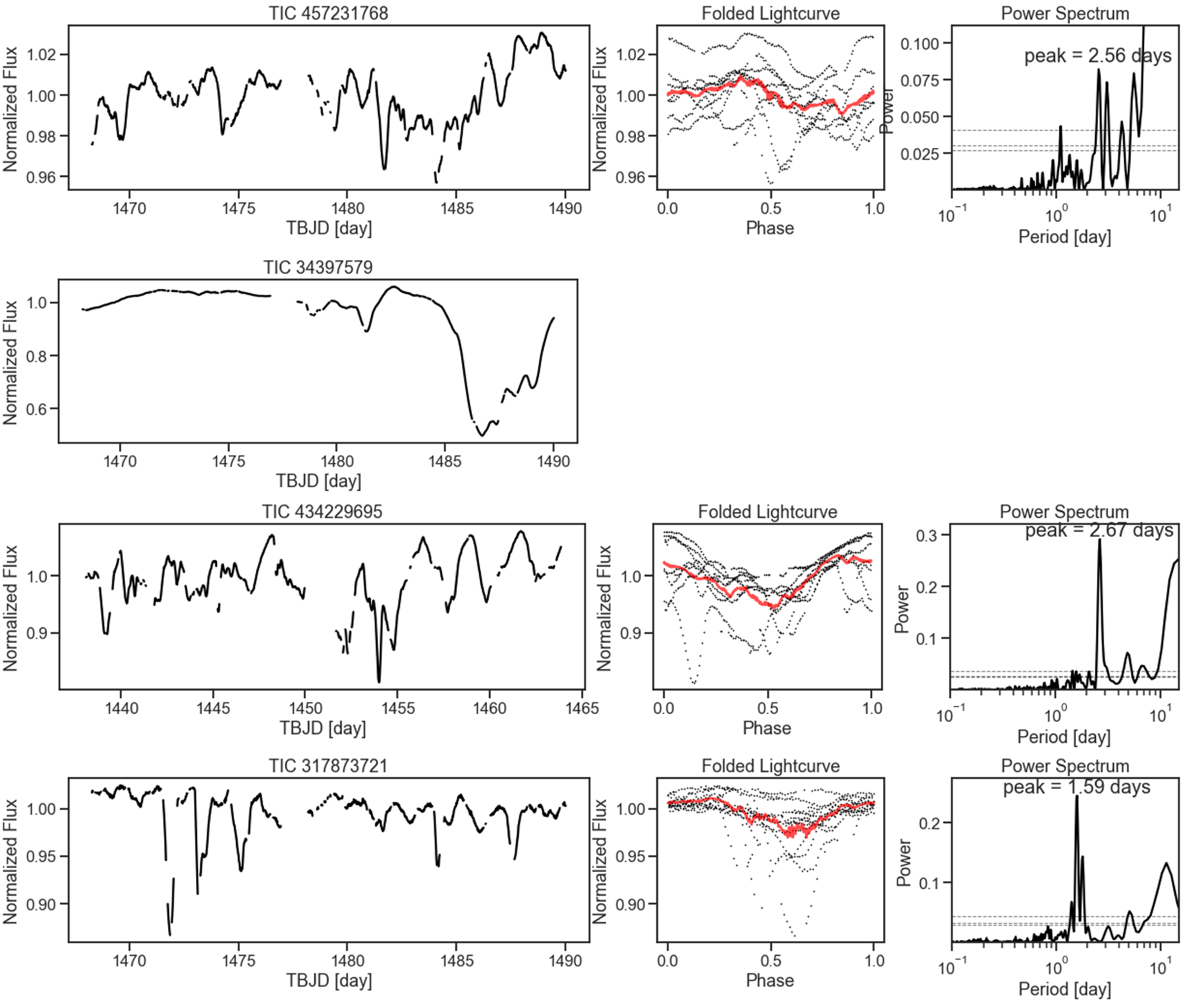}
  \caption{\textit{Left column}: TESS FFI light curves for observed targets.  \textit{Middle column}: The folded light curve for each object with the period corresponding to the peak of the power spectrum. The black dots represents all data, and the red line is the smoothed signal by convolving the box kernel. \textit{Right column}: Lomb-Scargle periodogram of the TESS light curve. Since the dip is appeared only once in the light curve of TIC 34397579 and we cannot measure the periodicity, we do not show the folded light curve and the power spectrum of it.}
  \label{fig:LCs}
\end{figure*}
\begin{figure}
  \centering
  \includegraphics[width=8.5cm]{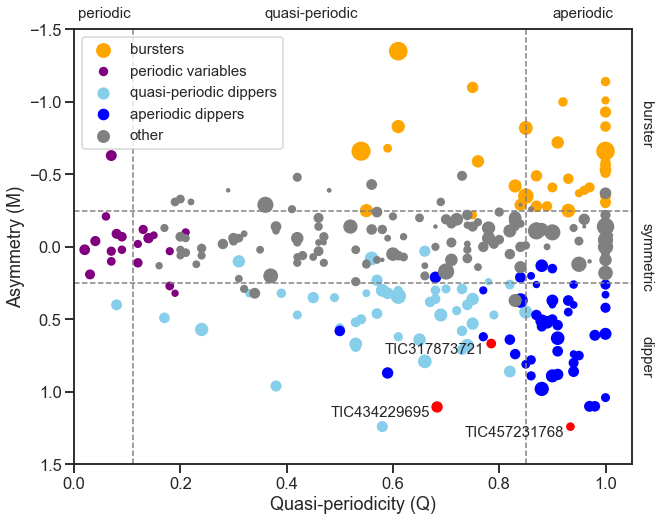}
  \caption{Classification of the light curve morphology, using the definitions of $Q$ and $M$ defined by \citet{Cody2014}. Targets of this paper are plotted with red circles compared to K2 light curves of variable stars in Upper Scorpius and $\rho$ Ophiucus star-forming regions as determined by  \citet{Cody2018}. Markers are scaled by the square root of the variability amplitude. The dashed lines divide the different classes suggested by \citet{Cody2018}. }
  \label{fig:QM}
\end{figure}
We reviewed the periodicity of the light curve for each target from the first-year data of the TESS mission and classified them according to the morphologies.
A set of statistical metrics was introduced to classify YSO light curve shapes into different categories by \citet{Cody2014} which were the flux asymmetry, ``M'', and the stochasticity, ``Q''.
If the amplitude of a light curve variation is symmetric, $M \in [ -\infty, \infty ]$ will be equal to zero, while $M>0$ means that the dimming time is relatively short.
$Q \in [0,1]$ will be equal to zero for a perfectly periodic light curve and will be close to 1 for a stochastic one.
We followed the classification from \citet{Cody2018}, which classified variable stars into some classes such as burster ($M<-0.25$), quiasi-periodic dipper ($0.15<Q<0.85$ and $M>0.25$), and aperiodic dipper ($Q>0.85$ and $M>0.25$).

Figure \ref{fig:LCs} shows light curves of the TESS FFI and the result of folding it by the period most related to the dip.
We searched for these periods by using Lomb-Scargle Periodogram.
Figure \ref{fig:QM} shows the variability types according to M and Q statistics.
We found that TICs 317873721 and 434229695 are quasi-periodic dippers with periods of \SI{2.67}{day} and \SI{1.59}{day}, respectively, and TIC 457231768 is an aperiodic dipper with a period of roughly \SI{2.56}{day}, \rev{while we cannot measure the periodicity for TIC~34397579 because the dip appears only once in the light curve.}

\subsection{Spectral energy distributions}
\label{sec:SED}
\begin{figure*}
    \centering
    \includegraphics[width=16cm]{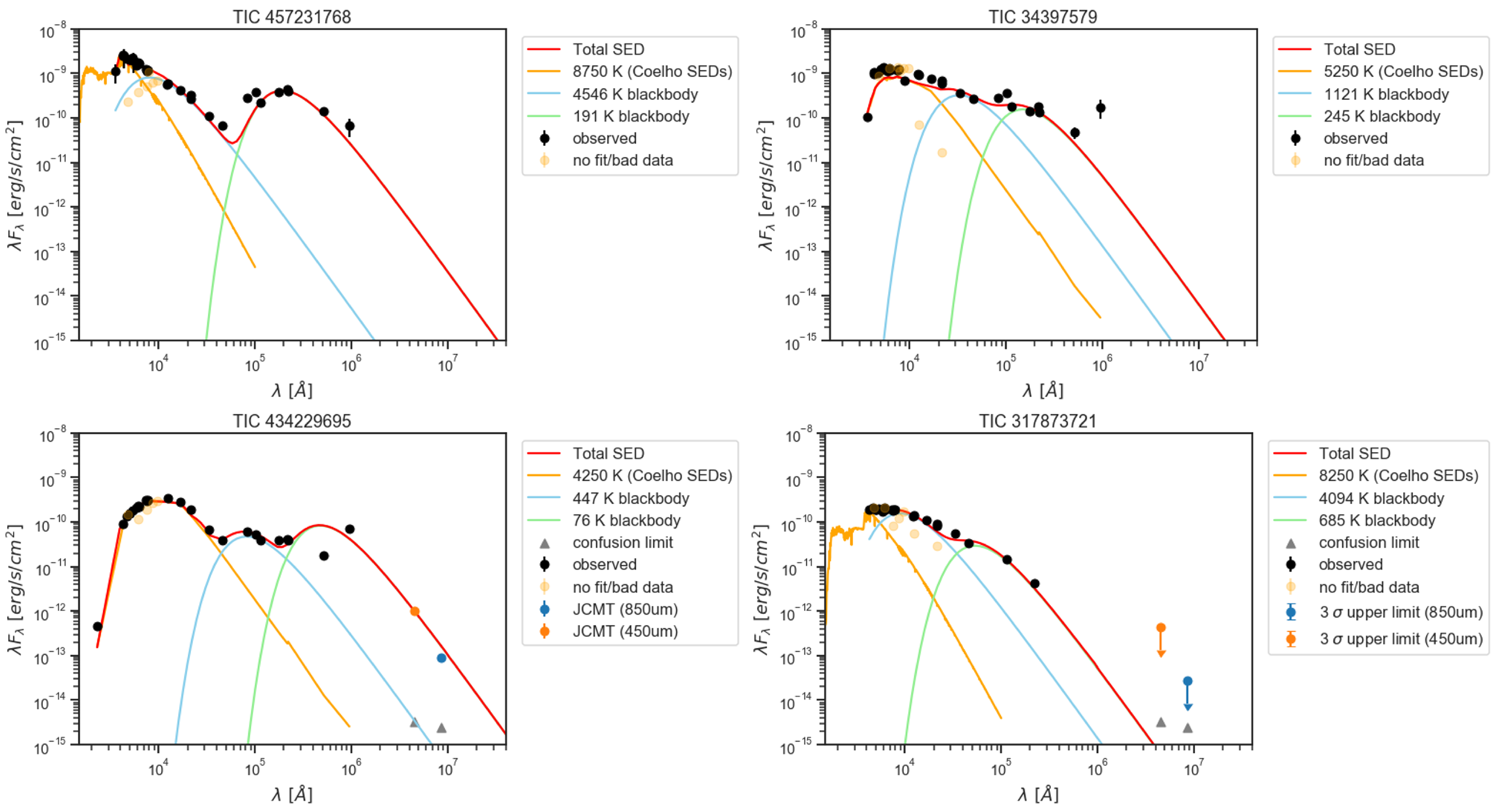}
    \caption{SEDs for each target built by using Virtual Observatory (VOSA). The orange lines are the best-fit theoretical model for data at wavelength below \SI{3.4}{\mu m} except for TIC 317873721. Since there might be a contribution from the companion star in the SED of TIC 317873721, we adopt the theoretical model with the effective temperature from TESS input catalog as the stellar component. The blue and green lines are blackbody radiations from multiple point sources, which are fitted by using the least-square method.  Two points at \SI{450}{\mu m} and \SI{850}{\mu m} of the SED for TIC 434229695 are obtained from our observation with JCMT. For TIC 317873721, we show $3\sigma$ upper limits.}
    \label{fig:SEDs}
\end{figure*}

We also revisited the spectral energy distributions (SEDs) of dippers we observed.
From one of our definitions of the dipper described in Paper I, they show infrared excess in the SEDs (Figure \ref{fig:SEDs}). 
We built the SEDs by using Virtual Observatory \citep[VOSA;][]{Bayo2008} and calculated the spectral index  
$\alpha\left(\equiv \frac{\Delta\left(\log \left(\lambda F_{\lambda}\right)\right)}{\Delta(\log (\lambda))}\right)$
 between \SI{3.4}{\mu m} and \SI{22}{\mu m}.
The results are between $-1.28$ and $-0.014$ (listed in Table \ref{tab:plausible}), which implies that they are in the phase of Class I\hspace{-1pt}I-I\hspace{-1pt}I\hspace{-1pt}I according to the discussion of classification of disks \rev{\citep{Lada1987}}.

To characterize their disks for more detail, we performed sub-millimeter observations on JCMT for two dippers whose memberships were not identified in Paper I (TIC~434229695 and TIC~317873721).
Unlike the other two objects, which are young and are thought to have disks, the nature of the disks of these objects was unknown.

TIC~434229695 was detected with 25~mJy at 850~$\mu$m with the signal-to-noise ratio of 17 and with 156~mJy at 450~$\mu$m with the signal-to-noise ratio of 6.9.  The detection of sub-mm emission indicates that the star is in a dusty environment.  The $F_{\nu}=$25~mJy of 850~$\mu$m emission corresponds to the total dust mass $M_d$ of $\sim 0.1~M_J$\rev{, as calculated using the formula in \citet{Hildebrand1983}, }
$M_d = d^2 F_{\nu}/\kappa_{\nu} B_{\nu}(T)$, where $T$ is the temperature, $d$ is the distance, $\kappa_{\nu}$ is the opacity per unit dust mass and $B_{\nu}(T)$ is the Planck function.
Here, we have assumed $T=20$~K and $\kappa=8.5$~g/cm$^2$ as in other sub-mm photometry studies \citep[e.g., ][]{Ansdell2016Lupus}.

\begin{figure}
    \centering
    \includegraphics[width=8cm]{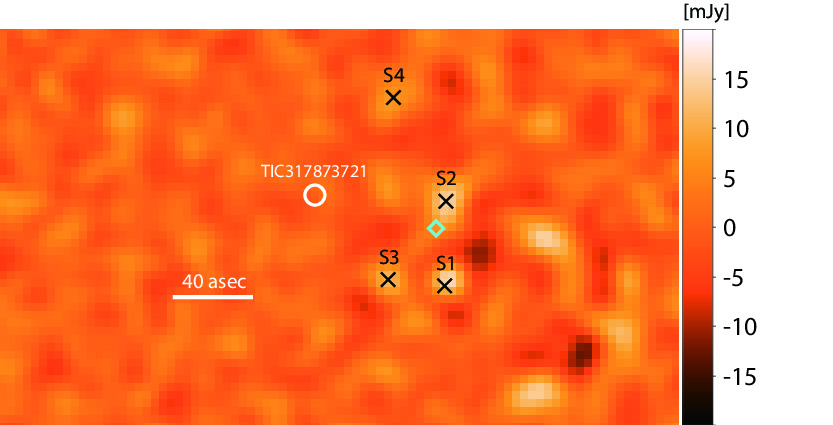}
    \caption{JCMT 850~$\mu$m map around TIC~317873721.  The position of the object is marked with white circle and the point sources S1-S4 identified by the pipeline are marked with black crosses.  The position of IRAS~05555-1405 is marked with the cyan diamond.}
    \label{fig:JCMT}
\end{figure}

TIC~317873721 was not detected with JCMT.  The 3$\sigma$ upper limit for TIC~317873721 is 67~mJy for 450~$\mu$m and 7.8~mJy for 850~$\mu$m.  However, the SCUBA-2 pipeline tentatively identified four weak point-like emission at $\sim$1~amin ($\sim 0.2$~pc) away from the object at 850~$\mu$m (Figure \ref{fig:JCMT} and Table \ref{tab:JCMTPointSource}).  Among the four, the source S1 is located 3~asec  away from 2MASS~J05574918-1406080 and S2 is located 6~asec away from 2MASS J05574947-1405336, both within the beam size ($\sim$13~asec) of JCMT.  The two 2MASS objects are both identified as YSO.  Moreover, IRAS~05555-1405 is located close to the sources.  The two 2MASS and one IRAS sources are considered to be associated with the cloud called VdB~64 \citep{Lee2009}.
We confirmed that the Gaia DR2 proper motions and parallaxes of the two 2MASS sources are close to those of TIC~317873721 (pm$\sim (-2,1.8)$\si{mas.yr^{-1}} and parallax$\sim$\SI{1.5}{mas}) and therefore we consider that TIC~317873721 is a member of VdB~64.

\begin{deluxetable}{ccccl}
\tabletypesize{\footnotesize}
\tablecaption{The list of point sources detected around TIC~317873721 \label{tab:JCMTPointSource}}
\tablehead{ \colhead{Source} & \colhead{RA} & \colhead{Dec} & \colhead{Peak Intensity (mJy)} 
}
\startdata
    S1 & 05:57:49.3 & -14:06:10.7 & 9.4  \\
    S2 & 05:57:49.3 & -14:05:28.4 & 9.2 \\
    S3 & 05:57:51.3 & -14:06:07.6 & 6.9  \\
    S4 & 05:57:51.1 & -14:04:36.6 & 4.6 \\
\enddata
\end{deluxetable}  



In Figure \ref{fig:SEDs}, we show the SEDs and the fitted model for each target.
For the stellar component ($\lambda < $\SI{3.4}{\mu m}) of the fitted model, we use the best-fit theoretical spectra by \citet{Coelho2014} obtained from VOSA, then for residual data, fit blackbody radiations from multiple point sources by using the least-square method.
Some points with bad quality for whatever reason were not used for the fitting.
The model of TIC 317873721 is not built in this way, since the infrared excess is thought to be related not only to the disk component but also to the companion star (Sec. \ref{sec:TIC317873721}).
Therefore, we set the theoretical model with $T_{\mathrm{eff}}=$\SI{8250}{K} as the primary star's component instead of the best fit model selected by VOSA.
\rev
{
Similarly, we considered that the points that were not used for the fit of TIC~457231768 were the variability due to the companion in this binary system (SB2), as reported in \citet{Doering2009} \citep[see also ][]{Alecian2013}. 
The temperatures of primary and secondary are 9000 K and 5000 K, respectively, which roughly match the temperatures of the fitted blackbody radiations.}

All fitted models were composed of three components.
For TICs 34397579 and 434229695, we can see the flux in the $10$--\SI{100}{\mu m} region does not decrease with a simple power law, but increases slightly with wavelength, and decreases with a steep slope in the sub-mm region.
Although these are not unique models to explain this feature, the flared disk model  \citep{Chiang1997} or the ``puffed-up'' inner rim  \citep{Isella2006} are candidate models to explain it.
The former is a two-layered disk with different temperatures on the surface and inside the disk.
Only half of the photon energy from the star absorbed by the disk surface is re-radiated, while the other half heats up the inside of the disk. 
As a result, in addition to the light from the star and the reflected light from the disk surface, longer wavelength light from inside the disk is also observed.
In the latter model, there is a region in the center of the disk where dust sublimates, and the entire edge of the inner disk expands due to the radiation from the star.

\subsection{Estimation of the stellar age from \ce{Li} absorption line}
\begin{deluxetable*}{ccccl}
\tabletypesize{\footnotesize}
\tablecaption{Results of Stellar Properties Measured from HDS Spectra\label{tab:result}}
\tablehead{
 & \multicolumn{2}{c}{\ce{H\alpha}} \vspace{-0.1cm} & \colhead{Li} &  \\ \cline{2-3}
 \colhead{TIC ID} & \colhead{W$_{10}$ (\si{km.s^{-1}})} & \colhead{M$_{acc}$ (M${_\odot}$\,yr$^{-1}$)} & \colhead{EW (\si{m\angstrom})} & \colhead{Instruments} 
}
\startdata
    457231768 & 507 & $1.07 \times 10^{-8}$ & $27 \pm 6$ & HDS\\
    34397579 & 549 & $2.78 \times 10^{-8}$ & $150 \pm 7$ & HDS\\
    434229695 & 360 / 555 & $4.08 \times 10^{-10}$ / $3.18 \times 10^{-8}$ & $254 \pm 9$ / $247 \pm 9$ & HDS, CAFÉ, SCUBA-2\\
    317873721 &  -- / 526 & -- / $1.65 \times 10^{-8}$ & -- & HDS, CAFÉ, SCUBA-2, HIDES \\
\enddata
\tablecomments{Results of \ce{H\alpha} and \ce{Li} of TICs 434229695 and 317873721 are derived from the spectra observed in 2019 / 2020. All other results are from spectra observed in 2019.}
\end{deluxetable*}  
\begin{figure}
    \centering
    \includegraphics[width=8.5cm]{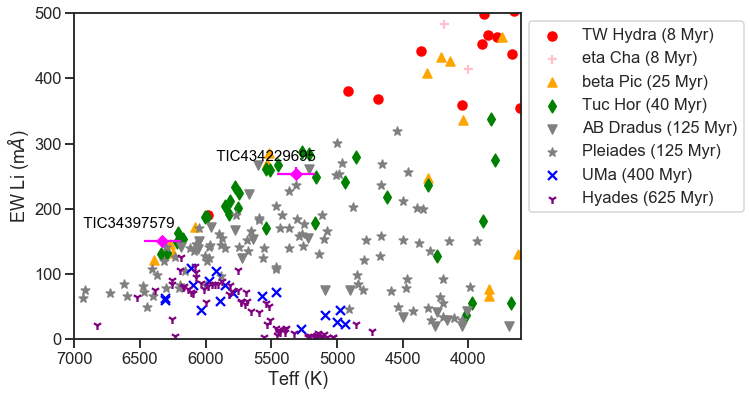}
    \caption{The EWs of \ce{Li} and the effective temperatures  \citep[c.f.][]{Gaidos2019}. Comparing these parameters of the dippers (plotted pink diamonds with error bars) to those of the nearby young star clusters, the ages for each target can be estimated. Because there exist less samples around $T_{\mathrm{eff}}>$\SI{7000}{K}, we exclude TIC 457231768 ($T_{\mathrm{eff}}=$\SI{8785}{K} from TICv8) from this figure.}
    \label{fig:EWLi}
\end{figure}
In the spectra except for TIC 317873721, we confirmed the \ce{Li} absorption line (\SI{6707}{\angstrom}), which \revrev{indicates they are in the pre-main sequence} because lithium depletion occurs rapidly after the core temperature becomes high enough ($\approx 3\times 10^{7}$\si{K}) to start burning lithium.
The fact that the \ce{Li} absorption line was not detected in the spectrum of TIC 317873721 can be interpreted that the high temperature of the star has already fused all Lithium.

We measured equivalent widths (EWs) of \ce{Li} of the spectra observed with HDS/Subaru in 2019 and 2020 by fitting the absorption line with the gaussian function.
Results are listed in Table \ref{tab:result}.
Following Figure 13 in \citet{Gaidos2019}, we plotted \ce{Li} EWs measured from the spectra taken in 2019 with that of the dippers with those of nearby young moving groups in Figure \ref{fig:EWLi}.
The effective temperatures are set to the values in \rev{the TESS Input Catalog version~8 \citep[TICv8; ][]{Stassun2019}}.
\rev{
From Figure \ref{fig:EWLi}, we roughly estimate their ages to be $<$ \SI{125}{Myr} according to the location in that parameter space, \revrev{in the same way as Eric Mamajek, who fitted polynomials to the data of clusters in this type of diagram}\footnote{http://www.pas.rochester.edu/~emamajek/images/li.jpg}.
Note that the ages estimated from \ce{Li} EWs may have large uncertainty and should only be seen as a rough estimate.
One of the reasons for this is that \revrev{we did not take into consideration the veiling, which causes lines to appear shallower due to excess emission from an accretion shock.} Since correcting this effect increases the EW, the age should not be higher than the estimated ages.
Another reason is that the compared data from each moving group itself show a large scatter, for example, due to the different achievable signal-to-noise ratios and the stellar properties such as rotational speed.
It means that a clear boundary of age in this figure is difficult to be determined.
}

\revrev{Indeed, other results estimate ages that are much younger than this upper limit}, for example, in Figure 4 of Paper I, which shows the Hertzsprung–Russell (HR) diagram of dippers constructed from the Gaia DR2 data and compares it with MESA isochrones, both dippers are plotted between lines of the ages of \SI{1}{Myr} and \SI{10}{Myr}.
Also, the age of TIC 34397579 reported in \citet{Arun2019} is $\sim$\SI{7}{Myr}.


\subsection{\ce{H\alpha} emission lines}
\rev{The \ce{H\alpha} emission line is one of the features of Classical T Tauri stars or weak-line T Tauri stars.}
The observed line profiles of the emission lines \revrev{of the Classical T Tauri stars} are well explained by the accretion model along the stellar magnetic field as follows; in young stars, gas is accreted from the edge of the inner disk to the stellar surface.
When the gas reaches both poles at a free-fall speed along the magnetic field, X-rays and EUV-rays are emitted by its shock.
Because the gas is optically thick, this radiation with a short wavelength is immediately absorbed and the visible and UV radiation is re-emitted \citep{Calvet1998,Gullbring2000}.
The red-shifted or blue-shifted absorptions are often seen in these emission lines, which is considered to be caused by disk wind or accretion flow \citep{Kurosawa2006}.

\begin{figure*}
    \centering
    \includegraphics[width=16.0cm]{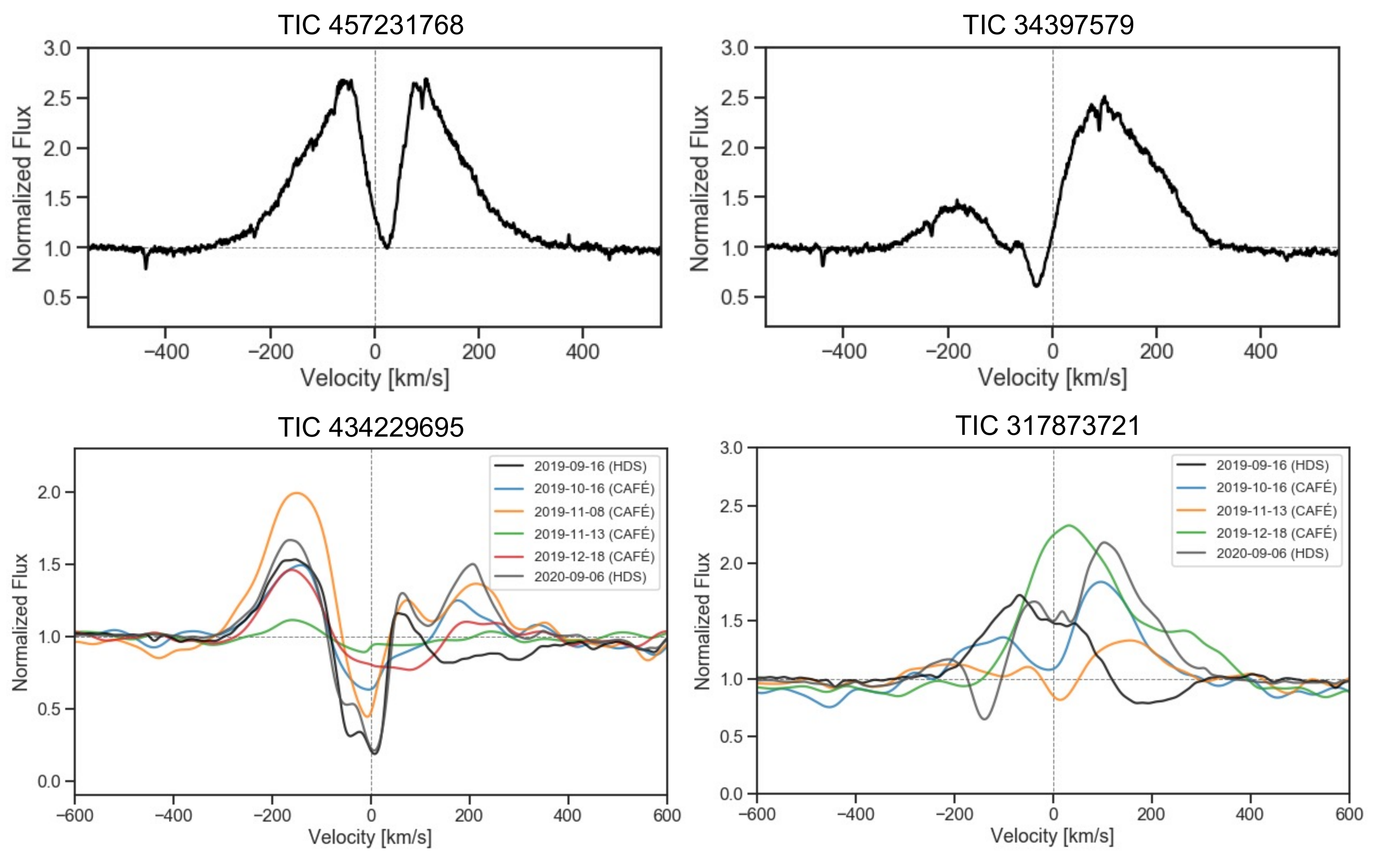}
    \caption{Observed spectrum around \ce{H\alpha} emission line. The vertical dashed line is at the rest wavelength (\SI{6562.8}{\angstrom}). To show the line profiles clearly, observed spectra of TICs 434229695 and 317873721 are smoothed by convolving a Gaussian function. See Table \ref{tab:result} for observational instruments.}
    \label{fig:Hals}
\end{figure*}

\rev{All of our targets show the \ce{H\alpha} emission line, and we found that, from the broad feature of the emission line, they are considered to be young stars with accretion.
According to \citet{White2003}, stars with the \ce{H\alpha} broader than \SI{270}{km.s^{-1}} at $10\,\%$ maximum intensity are classified as  accreting, and all the targets exceed this threshold (Figure \ref{fig:Hals}). } 
TICs 457231768 and 34397579 are known to show this emission line from the previous observations.
In this observation, a double-peaked line profile and a blue-shifted absorption line are observed in the spectrum of TICs 457231768 and 34397579, respectively. 
\rev{They will be described in Section \ref{sec:TIC457231768}, \ref{sec:TIC34397579}.}

We checked the line profile variations from additional observations for TICs 434229695 and 317873721.
The line profiles of TIC 434229695 vary and sometimes even disappear, but they are basically that the peak on the red-shift side is less than half of that on the blue-shift side. 
The variation of the line profile for TIC 317873721 is more complicated as shown in Figure \ref{fig:Hals}.
The interpretation of each emission-line profile will be dealt with in Section \ref{sec:scenarios}. 

Emission in the \ce{H\alpha} line can also arise from chromospherically active stars, instead of accretion onto the surface of a star from a disk. 
It can be distinguished whether objects are non-accreting or magnetospheric accreting by the line widths and line profiles. 
The large-velocity magnetospheric accretion columns produce broad ($>$\SI{200}{km.s^{-1}}) and asymmetric line profiles  \citep{Natta2004}. 
We measured the width at $10\%$ of the line peak (W$_{10}$) in Subaru/HDS spectra, which are listed in Table \ref{tab:result}. 
Then, we estimated the accretion rates of their disk according to the empirical relation obtained by \citet{Natta2004}, 
\begin{equation}
\begin{aligned} \log \left(M_{\mathrm{acc}}\left(M_{\odot} \mathrm{yr}^{-1}\right)\right)=&-12.89(\pm 0.3)+9.7(\pm 0.7) \\ & \times 10^{-3} \Delta V\left(\si{km.s^{-1}}\right). \end{aligned}
\end{equation}
The results of the mass accretion rate $\sim 10^{-8} M_{\odot} \rm yr^{-1}$ are consistent to the model prediction of observed line profiles by \citet{Kurosawa2006}.

\subsection{Radial velocity variation of TIC 317873721}
\label{sec:RV}
\begin{figure*}
    \centering
    \includegraphics[width=15.0cm]{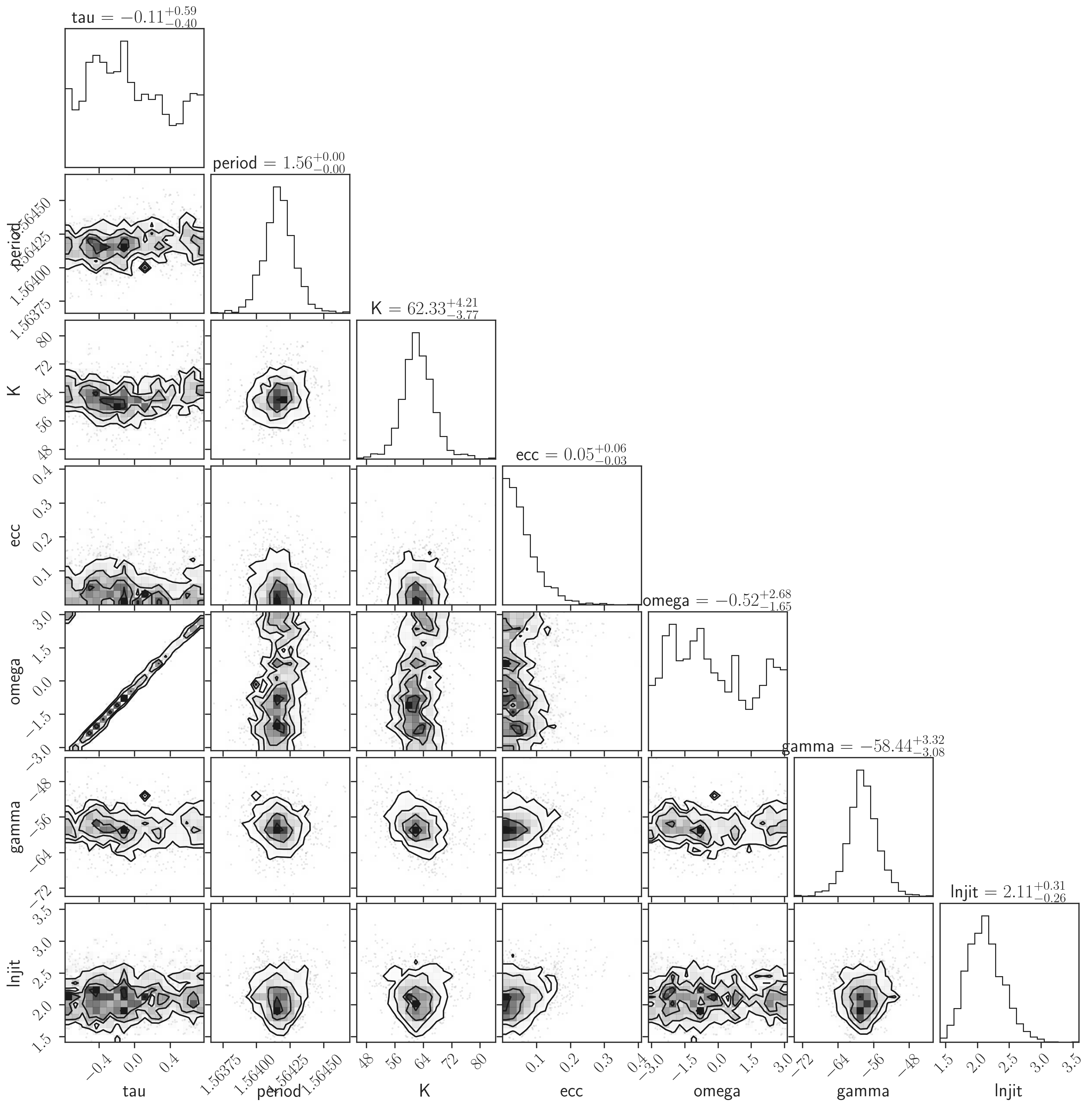}
    \caption{Posterior distribution of the RV model $(\tau [\si{day}], P_\mathrm{orb} [\si{day}], K [\si{km.s^{-1}}], e, \omega [\si{radian}], \gamma [\si{km.s^{-1}}], \ln \mathrm{jitter})$. The paramter of ``$\ln \mathrm{jitter}$'' is the natural log of the $\sigma_{\mathrm{jitter}}$ --- see text for more information.}
    \label{fig:MCMC}
\end{figure*}
\begin{figure}
    \centering
    \includegraphics[width=8.0cm]{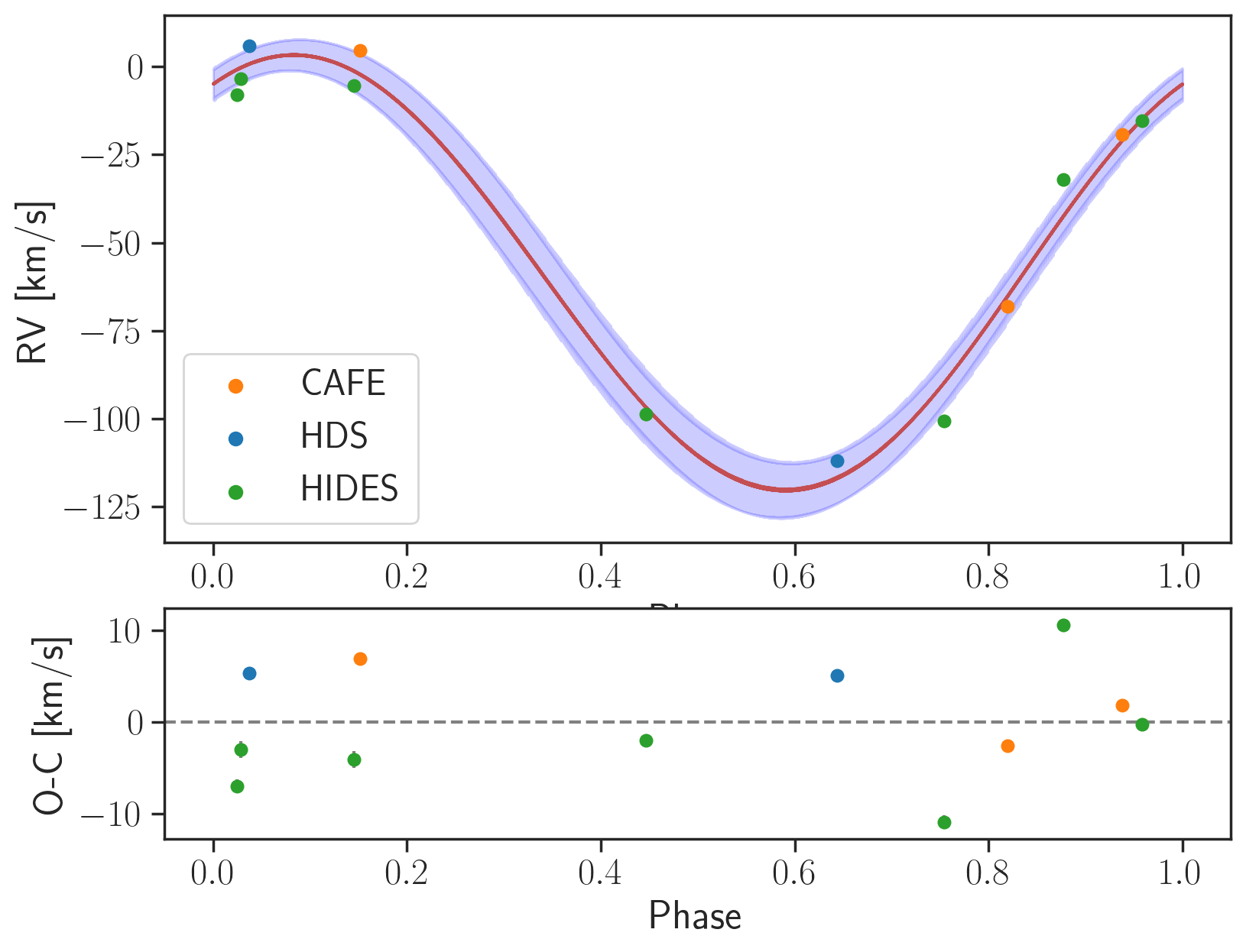}
    \caption{The mean (solid line) and standard deviation (shaded region) of the fitted models from MCMC sampling (top), and the difference from the fitted curve (bottom).}
    \label{fig:RV_foldedrv}
\end{figure}
We found that TIC 317873721 shows a large radial velocity variation which indicates there is a close-in object.
These radial velocities were derived by cross-correlation of the observed spectra with the template spectrum around Mg triplet ($5150$ -- \SI{5200}{\angstrom}).
We used the high-resolution synthetic spectra of $T_{\mathrm{eff}}=$ \SI{8200}{K} calculated by PHOENIX \citep{Husser2013} as the template spectrum. 
For the spectra with low S/N, we removed outliers by sigma-clipping and smoothed them by convolving a Gaussian function.
Errors were estimated by Monte-Carlo simulation with the mock spectrum for each observation, resulted in about $1$--$3\,\mathrm{km\,s^{-1}}.$

The radial velocity, $V(t)$, of the primary star in a binary system is given by
\begin{equation}
V(t)=\gamma+K_1\{\cos [\theta(t)+\omega]+e \cos \omega\},
\label{eq:rv}
\end{equation}
where $\gamma$ is the long term mean or systemic velocity of the binary, $\theta(t)$ is the true anomaly as a function of time $t$, $\omega$ is the argument of periastron, and $e$ is the eccentricity. 
$K_1$ is the radial velocity semi-amplitude of the primary star, which is given by
\begin{equation}
K_{1}=\frac{n a_{1} \sin i}{\sqrt{1-e^{2}}}
\end{equation}
where $n=2\pi / P_{\mathrm{orb}}$ is the mean motion, $P_{\mathrm{orb}}$ is the orbital period, and $a_1$ is the semi-major axis of the primary star’s orbit.
The velocity semi-amplitude of the secondary star in the binary, $K_2$, is defined in an analogous manner using $a_2$. 
$\theta(t)$ in the equation \ref{eq:rv} is related with time through the following two equations\rev{, as introduced in \citet{Green1985}}:
first, an expression between the true anomaly and the eccentric anomaly, $E$, and second, between the eccentric anomaly and time (the latter is known as Kepler’s Equation, where $\tau$ is the time of perihelion passage).
\begin{equation}
\tan \frac{\theta}{2}=\sqrt{\frac{1+e}{1-e}} \tan \frac{E}{2}
\end{equation}
\begin{equation}
t=\frac{E-e \sin E}{n}+\tau
\end{equation}
We searched the orbital parameters of the primary star by using Markov chain Monte Carlo (MCMC) sampling.
The choice of prior distributions for each parameters is given in Table \ref{tab:prior}.
In the MCMC fitting, we assume that the measurement errors for RVs to $\sigma = \sqrt{\sigma_{i}^2 + \sigma_{\mathrm{jitter}}^2}$, where $\sigma_{i}$ is an internal error of the $i$th data point and $\sigma_{\mathrm{jitter}}$ is any other excess scatter that is not included in $\sigma_{i}$ \citep[c.f. ][]{Masuda2021}.
Figure \ref{fig:MCMC} shows the result of this sampling, and the folded radial velocities and the mean of the fitted models are shown in Figure \ref{fig:RV_foldedrv}.
Because the data is sparse in terms of the phase, the parameters of $\tau$ and $\omega$ were not well determined as shown in the near linear relationship in Figure \ref{fig:MCMC}. 
If we collect several data with various phases from further observations, all the parameters should be well determined.
\begin{deluxetable}{ccccl}
\tabletypesize{\footnotesize}
\tablecaption{The assumed prior distribution of orbital parameters\label{tab:prior}}
\tablehead{
\colhead{Parameter} & \colhead{Prior} & \colhead{Min} & \colhead{Max} 
}
\startdata
    $\log K$ & Uniform & -2 & 2 \\
    $e$ & Uniform & 0 & 1 \\
    $\cos \omega$ & Uniform & -1 & 1 \\
    $\sin \omega$ & Uniform & -1 & 1 \\
    $P_{\mathrm{orb}}$ & Uniform & $1.56\times0.99$ & $1.56\times1.01$ \\
    $\tau$ & Uniform & $-1.56/2$ & $1.56/2$ \\
    $\ln jit$ & Uniform & -5 & 5 \\
\enddata
\end{deluxetable}  

Then, we estimated the mass of the companion.
 The binary mass function $f$ is given by
\begin{equation}
    f=\frac{M_{2}^{3} \sin ^{3} i}{\left(M_{1}+M_{2}\right)^{2}}=\frac{P_{\mathrm{orb}} K^{3}}{2 \pi G} (1-e^{2})^{3/2},
\end{equation}
where $M_1$ and $M_2$ are the stellar masses and $G$ is the gravitational constant.
Given the mass of TIC 317873721 is $M_1=2.01 M_{\odot}$ from TICv8, the minimum mass of the unseen object is $M_2=0.65 M_{\odot}$ which is calculated by setting $i=90^{\circ}$.
In addition, the semi-major axis of this binary system can be estimated by Kepler's third raw,
\begin{equation}
    \frac{a^3}{P_{\mathrm{orb}}^2} = \frac{G}{4\pi^2}(M_1 + M_2),
\end{equation}
and is $a\sim$ \SI{0.036}{au} $=7.9 R_{\odot}$.

The spectral feature from the unseen companion in high resolution spectra could not be found.
If the \ce{Li} absorption line was in the spectra, it could sign the existence of a cooler companion, but it was not detected. 
\rev{
We also tried to derive the radial velocities of the secondary by the cross-correlation of the HDS spectra after removing the feature of the primary star.
We searched for signs of the secondary in the spectrum in the following process;
we firstly smoothed the raw spectrum by taking the median filter with the rotation speed of the primary star, $v \sin i \sim 40$ \si{km.s^{-1}}.
This rotation speed was roughly estimated by the cross-correlation with the template spectrum of a similar temperature star. 
Then, we divided the raw spectra by the smoothed ones, and because the smoothed spectrum can be regarded as the spectrum of the primary star, there should remain the features from the companion.
Finally, we took the cross-correlation of the remaining spectrum and the stellar template spectrum, but no signals could be detected.
}
It might because the contrast ratio of the primary star and the companion star is not enough to be detected at such noise level.
For example, when we set the effective temperature and the radius of the companion to be $T_\mathrm{eff}=$\SI{4100}{K} and $R=0.7R_{\odot}$, the contrast ratio at \SI{5200}{\angstrom} become $\sim 6 \times 10^{-3}$.

\section{Plausible scenarios for the four dippers}
\label{sec:scenarios}
The plausible scenarios of the dimming mechanisms for the four dippers are summarized in Figure \ref{fig:dipper_scenarios} and Table \ref{tab:plausible}.

\subsection{Dippers around \rev{the Orion complex}}
In this section, we discuss the dimming mechanisms for TIC 457231768 (HD 29409) and TIC 34397579 (V 1650 Ori).
\rev{The Orion molecular cloud complex (or, simply, the Orion complex) is one of the largest star-forming regions near the Sun.}
These targets belong to Orion C and D, respectively, which are older parts of the Orion complex with little molecular gas  \citep{Kounkel2018}.
The most likely cause of aperiodic and quasi-periodic dimming of these dippers is ``dusty disk wind.''
It is a scenario in which the dust is lifted by the disk wind generated by the XUV radiation from the star and the magnetic field, and the dust blocks the light in the line of sight, resulting in dimming.
The absorption line seen in \ce{H\alpha} emissions of these targets indicates the outflow from the disk.
The detailed properties of each object are described in the following subsections.

\subsubsection{TIC 457231768}
\label{sec:TIC457231768}
TIC 457231768 was previously estimated about its age as $\sim$\SI{4.8}{Myr} in Paper I from the catalog of  \citet{Kounkel2019}, but \citet{Arun2019} reported that it is a Herbig Be star with the age of $\sim$\SI{7}{Myr}.
Also, \citet{Alecian2013} found that it is a spectroscopic binary (SB2).

We confirmed that the \ce{H\alpha} emission line with the slightly red-shifted absorption was seen in the spectrum (Figure \ref{fig:Hals}).
\rev{Compared to the past observations, the line profile is likely to show variability. 
In \citet{Torres1995}, it is reported that the emission has a slightly stronger blue-shifted peak than the red-shifted one. 
But in \citet{Vieira2003}, it is reported to be a double-peaked shape just like our observation.
These line profiles suggest that this system is viewed from a high inclination angle ($i\sim 80^{\circ}$) \citep{Kurosawa2006}.}

\rev{Although the observed red-shifted absorption suggests the mass accretion, we considered that the cause of the dips was dust lifted above the disk midplane by the disk winds because this star is hot, dust would not be expected to survive in an accretion disk warp close to the star.
The inclination angle is enough high to observe the disk wind, which is known to be generated in a region $\gtrsim 30^{\circ}$ \citep{Blandford1982}.
}
For example, HD 163296, one of the dippers of Herbig Ae stars, was reported that dust clouds in its disk caused the variations of the brightness in the visible and near-infrared by \citet{Ellerbroek2014}.
Similar to TIC 457231768, the double-peaked \ce{H\alpha} emission line of HD 163296 was observed in the spectrum \citep{Acke2005}
, and the age was $\SI{6.52}{Myr}$. 
\rev{Therefore, we concluded that the more likely scenario to explain the dips for this star is the dusty disk wind scenario (Figure \ref{fig:dipper_scenarios} (a)).}

We calculated the size of dust clouds in the disk from the duration of the dimming.
\rev{In the calculation, we set \revrev{the location of the dust} for TIC~457231768 to about \SI{0.4}{au} from the star, adopting the estimate of the dust location causing the dipper phenomenon in HD~163296 made by \citet{Rich2020}.}
This is a reasonable value in the point that the sublimation radius of dust, which related to the dust sublimation temperature with
$R_{\mathrm{sub}} = R_{\star}(\frac{T_{\mathrm{sub}}}{T_{\star}})^{-2.085}$,
is $R_{\mathrm{sub}} \simeq 78R_{\odot}=0.36$\si{au} for TIC 457231768 when $T_{\mathrm{sub}}=1500$\si{K}.
With the dimming duration of $\approx 5$\si{days}, we estimated that the source of the dipper event has the azimuthal extent of \SI{0.2}{au}.

\subsubsection{TIC 34397579}
\label{sec:TIC34397579}
TIC 34397579 is a variable star of Orion type with a spectral type of F7.
In Paper I, it was classified as a member of Ori D south-1 ($\sim$\SI{3.0}{Myr}) according to the catalog of  \citet{Kounkel2019}.
In the TESS FFI light curve, there is a single large dip that lasts for about 4 days. 
Besides TESS FFI, quasi-periodic dimming was observed from the observation of ASAS-SN \citep{Pojmanski2002} and KELT \citep{Oelkers2018}.

In the observed spectrum, we confirmed \ce{H\alpha} emission line with the blue-shifted absorption (Figure \ref{fig:Hals}).
This absorption is caused by materials moving towards the observer, and the line profile is considered to be observed in a star with 
\rev{a fast wind observed at moderate inclination and }
a high accretion rate from the simulation by  \citet{Kurosawa2006}.
\rev{Compared to the past observations, it is not clear that whether there is an apparent change in the line profile.
Only in \citet{Rojas2008}, they reported it as ``complex emission profiles.''
In any case, the observed \ce{H\alpha} line profile may suggest the ``dusty disc wind'' scenario for the dipper phenomenon of TIC~34397579.}
Also, crystalline silicate features have been confirmed in the spectrum by \citet{Chen2016} from the observation with the Spitzer Infrared Spectrograph.
Protoplanetary disks contain crystalline dust grains where the dust temperatures are lower than the threshold value for their formation through thermal annealing of amorphous interstellar silicates.
The ``dusty disc wind'' scenario could explain emission from silicates even in the low dust temperature region due to the transport of particles by the disk wind \citep{Giacalone2019}.

\subsection{Dippers Showing \ce{H\alpha} Emission Variability}
We discuss the dimming mechanisms from our spectroscopic observations for TICs 434229695 and 317873721 in this section.
They are far from nearby molecular clouds and do not belong to any young moving groups and associations as shown in Paper I.
There is less spectroscopic observation for these targets, so this is the first time to characterize them in detail.
We observed them several times and found the variability of the \ce{H\alpha} emission line, which indicates the accretion flow or the interaction to an unseen object, as described below.

\subsubsection{TIC 434229695}
We found that the \ce{H\alpha} emission line of this target varies as shown in Figure \ref{fig:Hals}.
These line profiles, which have a double-peaked shape with a low peak on the \rev{red-shift} side overall, correspond to the model by  \citet{Kurosawa2006} with viewing the accretion flow onto the star from a high inclination angle.
In the model of \citet{Kurosawa2006}, it is also shown that the larger the inclination, the deeper the absorption feature.

\rev{To confirm the location of the source which cause dips, we calculated the Keplerian rotation radius $R_{K} = (GM_\star (\frac{P_{dip}}{2\pi})^2))^{1/3}$ that corresponds to the period of dips in the light curve.
Since $P_{dip}=2.56$\,days (from section \ref{sec:LCs}) and $M_{\star}=0.915M_{\odot}$ (from TICv8) for this target, then $R_{K}$ is $ 0.037$\,au.
This is consistent with a typical truncation radius $R_{t}$ of a T Tauri star \citep[$3R_{\star} - 5R_{\star}$][]{Shu1994}, about $0.027 - 0.045$\,au for TIC~434229695, at which the magnetic field truncates the inner disk and drives accretion flow through funnel flows.
}

Therefore, we concluded that \rev{the more likely scenario for the dipper phenomenon of this star} is the accretion flow generated along the axis of the magnetic field which is inclined to the axis of rotation, and that the viewing angle of the accretion flow changes with rotation, causing the quasi-periodic dimming and the variation of the line profile (Figure \ref{fig:dipper_scenarios} (b)).
This dimming mechanism is also applied to AA Tau, one of the typical dippers, and considered to be the typical scenario  \citep{Bouvier1999,Bouvier2014}.
\rev{If the simultaneous observation of spectroscopy and photometry is performed in the future, we will be able to confirm this scenario by connecting each line profile of the \ce{H\alpha} with the dimming phase.}
On the other hand, the unique point about this dipper is that it is located far from star-forming regions.
The estimated formation process for such objects is that they are formed in a small region with high-density gas and dust like ``Bok globule'', or they experienced the rapid dissipation of a surrounding molecular cloud.
It is known that \rev{dippers represent $\sim 20\%$ of Young Stellar Objects (YSOs) \citep[up to about 30-40\%, ][]{Alencar2010,Cody2014,McGinnis2015,Bodman2017,Cody2018}}, and TIC 434229695 suggests that there are many YSOs outside star-forming regions.

\subsubsection{TIC 317873721}
\label{sec:TIC317873721}
Although this star is located far from the Orion molecular cloud, it is considered to be a member of a star-forming region called VdB 64 (see Section \ref{sec:SED}).

From our observational results, we found that TIC 317873721 is a single-lined spectroscopic binary (SB1) with a circumstellar disk.
Interestingly, the estimated orbital period of this binary system ($P_{\mathrm{orb}}=$\SI{1.56}{days}) is close to the period of the dips seen in TESS light curve ($P=$\SI{1.59}{days}), so the dipper phenomenon seems to be strongly related to the binary motion.
Another intriguing feature of this star is the variability of \ce{H\alpha} emission line (Figure \ref{fig:Hals}) indicating ongoing accretion.
The circumbinary accretion onto binary is caused by the complex accretion streams;
the two circumstellar disks surrounding each star and the circumbinary disk are connected and all exchange mass via accretion streams launched at the inner edge of the circumbinary disk, as shown in numerical simulations \citep[e.g. ][]{Gunther2002,Kaigorodov2010,Munoz2016}.

Past observations of the T Tauri spectroscopic binary have revealed that some binaries show quasi-periodic photometric oscillations occurring at the binary orbital period, known as ``pulsed accretion'' \citep{Jensen2007,Muzerolle2013,Bary2014}.
In addition, they can show periodic variations in spectral veiling and emission-line intensities with orbital phase.
Such brightening is thought to be caused by the complex accretion streams as the material flowing from the circumbinary disk shocks when it collides with the circumstellar disk(s) or accretes onto the stellar surface(s) with the changing accretion rate \citep{Munoz2016}.

\begin{figure*}[t]
    \centering
    \includegraphics[width=16.0cm]{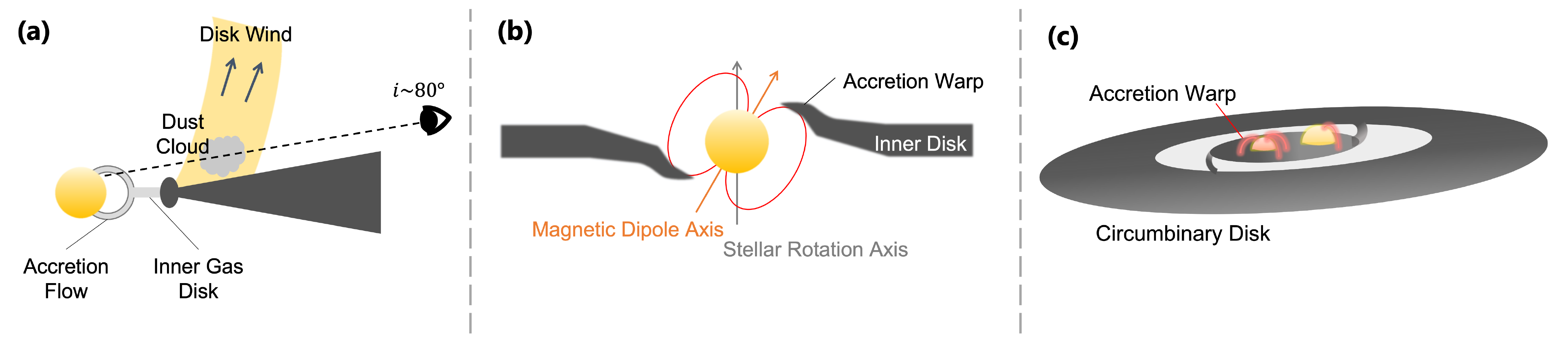}
    \caption{The sketches based on the plausible scenarios for dippers introduced in this paper. The key properties of each targets corresponding to the dimming mechanism are summarized in Table \ref{tab:plausible}. \rev{(c) Since TIC~317873721 is in a close-in binary system, the circumstellar dust structures may coalesce around both stars. }}
    \label{fig:dipper_scenarios}
\end{figure*}

The light curve of TIC 317873721 is not ``brightening'' periodically like pulsed accretion, but the depth of dimming is changing.
But we simply assumed that this system has the same geometry as a binary system with pulsed accretion, and that in this system, dust in the accretion stream, which rotates with the binary, hides the starlight and is observed as such dip \rev{when it is observed from a high inclination angle} (Figure \ref{fig:dipper_scenarios} (c)).
\rev{From our observation results, no clear periodicity of either the line profile and EWs of the \ce{H\alpha} emission with respect to the orbital phase of the binary was observed partially due to the low S/N data taken by CAHA/CAFÉ and Okayama/HIDES for this target.}
Further high S/N spectroscopic observations will be able to reveal the accretion in more detail.
Also, we plan to perform photometric observation in the near-infrared.
If the dip is due to interstellar dust, the effect of dust fading in K-band become smaller than in V-band \citep[e.g.][]{Grinin2018}. 
Therefore, the fading due to Eclipsing Binary (if any) may be observed in K-band observation and it will be able to confirm relations between the binary motion and the accretion stream.

\begin{deluxetable*}{clclccl}
\tabletypesize{\footnotesize}
\tablewidth{0pt}
\tablecaption{Summary of the Properties of Four Dippers and the Plausible Dimming Scenarios\label{tab:plausible}}
\tablehead{
 &  &  &  & \colhead{\ce{H\alpha}} \vspace{-0.4cm} & \colhead{RV} &  \\ 
 \colhead{TIC ID} & \colhead{Spectral Index} & \colhead{Disk Type$^{a}$} & \colhead{\ce{H\alpha} Line Profile$^{b}$} &  \colhead{}  \vspace{-0.4cm}& \colhead{} & \colhead{Plausible Dimming Scenario} \\
 &  &  &  & \colhead{Variability} & \colhead{Variation} &  
}
\startdata
457231768 & $\alpha = -0.014$;  & F         & double peak;          &   \rev{\checkmark (?)$^{d}$}  &              &  (a) disk wind                          \\
 & Class I\hspace{-1pt}I & & Disk winds \& High inclination & & &\\
34397579  & $\alpha = -0.418$;    & F         & I\hspace{-1pt}I\hspace{-1pt}I-B$^{c}$;                &     \rev{?$^{e}$}   &              &  (a) disk wind                         \\
 & Class I\hspace{-1pt}I & & Blue-shifted absorption due to disk wind & & &\\
434229695 & $\alpha = -0.111$;    & F         & I\hspace{-1pt}I-R$^{c}$;                 &    \checkmark       &              &  (b) accretion warp                          \\
 & Class I\hspace{-1pt}I & & Red-shifted absorption due to accretion  & & &\\
317873721 & $\alpha = -1.284$;    & F         & stochastic;           &    \checkmark       &  \checkmark  &  (c) binary \& accretion warp    \\
 & Class I\hspace{-1pt}I & & Interaction with binary motion? & & & 
\enddata
\tablecomments{$^a$ Disk type determined from Figure 3 in Paper I: F = full, $^b$ Observed line profiles and the estimated cause of them, $^c$ The classification of \ce{H\alpha} emission line profiles defined by \citet{Reipurth1996}, $^d$ In \citet{Torres1995}, the emission has a slightly stronger blue-shifted peak than the red-shifted one., $^e$ \citet{Rojas2008} reported as ``complex emission plofiles.''}
\end{deluxetable*}

\section{Discussion and Summary}
\label{sec:summary}
\rev{The dipper phenomenon, or the episodic/quasi-periodic dimming seen in the light curves, is considered to be related to the dust in the protoplanetary disk, which is lifted above the disk midplane by the disk wind, or by the accretion flow along the stellar magnetic field from the disk to the stellar surface.
Therefore, the observations of dippers can reveal the region close to the star in the protoplanetary disks.
While it is difficult to spatially resolve that region, high dispersion spectroscopic observation can give a clue to understanding such detailed properties of dippers.}

In this paper, we characterized \rev{four} newly found dippers by performing the follow-up observations.
These targets are selected from our catalog that contains a sizable, unbiased and homogeneous sample of dippers.
\rev{The observed spectra provide information on the presence of \ce{Li} absorption lines (for TICs 457231768, 34397579, and 434229695), an indicator of stellar youth, and \ce{H\alpha} emission lines (for all targets), suggesting accretion onto the star, as well as radial velocity variation caused by close-in binary (for TIC 317873721).
From these results, the dipper phenomena of our targets were considered to be caused by various mechanisms such as ``dusty disk wind'' (TICs 457231768 and 34397579) or dust in an accretion disk warp (TIC 434229695).
The accretion disk warp may also exist around TIC 317873721, a binary dipper, and rotate with binary based on the fact that the dimming period and the orbital period of the binary are almost the same. }

From the detection of the \ce{H\alpha} emission line, our observed dippers are still accreting even one of them (TIC 434229695) is located far from nearby molecular clouds.
By considering that dippers are common among YSOs \rev{\citep[$\sim 20 \% - 40 \%$:][]{Alencar2010,Cody2014,McGinnis2015,Bodman2017,Cody2018}}, there could be many YSOs out of star-forming regions.
In the point that they are less affected from the interstellar medium, our targets are suitable for the research of evolutional processes of the pre-main sequence stars by further follow-up observation.
Also, the measurements of radial velocities revealed a new aspect of a dipper with a circumbinary disk.
Dippers in binary systems have rarely investigated yet, TIC 317873721 become an important sample of understand a dipper mechanism.

\acknowledgments
We thank the anonymous reviewer for the helpful comments and suggestions.
We also thank Masayuki Tanaka, Ryotaroh Ishikawa, Takaharu Shishido, Raiga Kashiwagi, Suzuka Nakano and Takaho Masai for their help in preparing and carrying out the observation on September 16, 2019.
These data were acquired as a part of a practical training of observation with the Subaru telescope by SOKENDAI.

This study was supported by JSPS KAKENHI grant nos. JP18H04577, JP18H01247 (H.K.), JP20H00170, 21H04998 (T.K. and H.K.), 17H01103, 18H05441 (T.M. and M.M.), 19K03932  (T.M.), JP21K13965 and JP21H00053 (K.H.). In addition, this study was also supported by the JSPS Core-to-Core Program Planet2 and SATELLITE Research from the Astrobiology center (AB022006).

This research is based [in part] on data collected at Subaru Telescope, which is operated by the National Astronomical Observatory of Japan. We are honored and grateful for the opportunity of observing the Universe from Maunakea, which has the cultural, historical and natural significance in Hawaii.

The JCMT SCUBA-2 data are collected under program ID M20BP004.  The James Clerk Maxwell Telescope is operated by the East Asian Observatory on behalf of The National Astronomical Observatory of Japan; Academia Sinica Institute of Astronomy and Astrophysics; the Korea Astronomy and Space Science Institute; Center for Astronomical Mega-Science (as well as the National Key R\&D Program of China with No. 2017YFA0402700). Additional funding support is provided by the Science and Technology Facilities Council of the United Kingdom and participating universities and organizations in the United Kingdom and Canada.  Additional funds for the construction of SCUBA-2 were provided by the Canada Foundation for Innovation. 

\vspace{5mm}

\facilities{TESS, Subaru (HDS), CAHA (CAFÉ), Okayama; 1.88 m (HIDES), JCMT (SCUBA-2)}

\software{Astropy \citep{astropy:2013,astropy:2018}, astroquery \citep{Ginsburg2019}, matplotlib \citep{Hunter2007}, numpy \citep{Harris2020}, scipy \citep{Virtanen2020}, pandas \citep{Reback2020}, seaborn \citep{Waskom2021}, barycorrpy \citep{Kanodia2018},
corner \citep{corner}, JAX \citep{jax2018github}, NumPyro \citep{bingham2018pyro, phan2019composable}}

\bibliography{dipper}{}
\bibliographystyle{aasjournal}

\end{document}